Preliminary Thesis on the First Part of the ALLEGRO CFD Benchmark Exercise

ALLEGRO CFD BENCHMARK

PART 1

Flow Straightener Benchmark Description

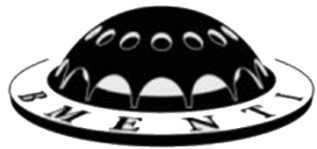

BUDAPEST UNEIVERSITY OF TECHNOLOGY AND ECONOMICS (BME)

Institute of Nuclear Technics (NTI)

2022

| Name | Occupation | Institution | Contact |
|---|---|---|---|
| **Gergely Imre Orosz** | **PhD student** | **BME NTI** | **orosz@reak.bme.hu** |
| Mathias Peiretti | MSc student | BME NTI/ Politecnico di Torino | matthias.peiretti@edu.bme.hu |
| Boglárka Magyar | BSc student | BME NTI | bogi0614@gmail.com |
| Dániel Szerbák | MSc student | BME NTI | szerbakdani@gmail.com |
| Dániel Kacz | PhD student | BME NTI | kacz.daniel@reak.bme.hu |
| Béla Kiss | Research assistant | BME NTI | kiss@reak.bme.hu |
| Gábor Zsíros | Research assistant | BME NTI | zsiros@reak.bme.hu |
| Prof. Attila Aszódi | Professor | BME NTI | aszodi@reak.bme.hu |

Budapest, Hungary, 2022.01.14.

## 1. Short description

At BME (Budapesti Műszaki és Gazdaságtudományi Egyetem - Budapest University of Technology and Economics) NTI (Nukleáris Technikai Intézet - Institute of Nuclear Technics), a 7 pin ALLEGRO rod bundle test section has been built in order to investigate the hydraulic behavior of the coolant in such design and to develop CFD models that could properly simulate the flow conditions in the ALLEGRO core. PIROUETTE (PIv ROd bUndlE Test faciliTy at bmE) is a test facility, which was designed to investigate the emerging flow conditions in various nuclear fuel assembly rod bundles. The measurement method is based on Particle Image Velocimetry (PIV) with Matching of Index of Refractory (MIR) method. Despite the working fluid in ALLEGRO will be helium, in this case water is in the same Reynolds number range of the ALLEGRO rod bundle. In the test loop, it was necessary to install a flow straightener that was able to condition the velocity field before the rod bundle. The results of CFD simulations could be used to improve the understanding of the inlet conditions in the rod bundle test section.

The herein proposed benchmark deals with the 3D CFD modeling of the velocity field within the flow straightener used in this test section The geometry of the flow straightener will be given to the participants in an easy-to-use 3D format (.tin, .stp or .stl for example). Link for ancillary files: https://drive.google.com/drive/folders/1Wz3oP-ug4dT25B6ZI0WwgpCKIdbSabi_?usp=sharing

In Figure 1, an overview of the test loop is provided. It can be seen where the flow straightener is placed in order to eliminate the effect of the elbow on the fluid flow.

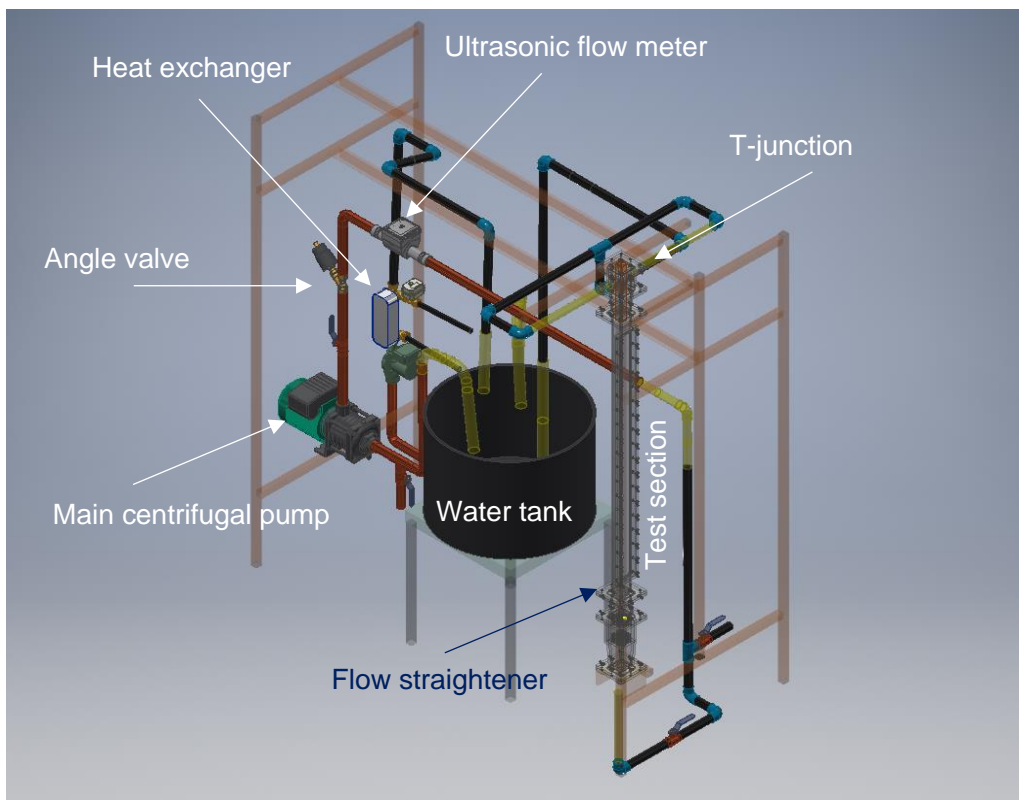

*Figure 1: The structure of the PIROUETTE facility*

## 1.1. Test section

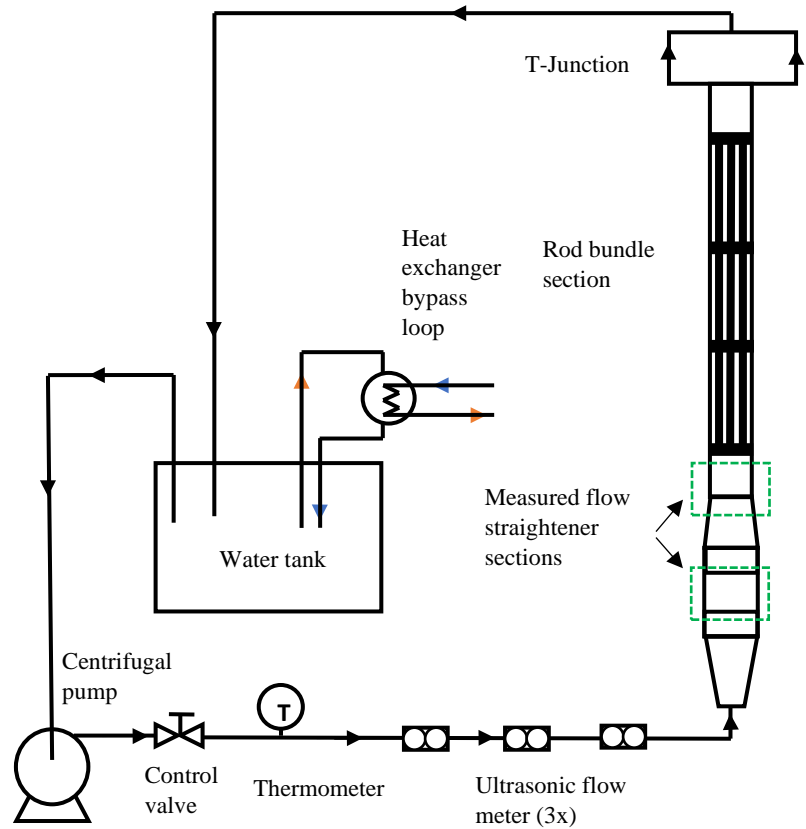

*Figure 2: The schematic of the PIROUETTE facility*

The structure and the main parts of the test facility can be seen in Figure 1. The installation contains a 1 meter long vertically arranged seven pin rod bundle in the test section. The water flow is provided by the main centrifugal pump (Type: Wilo MHIL 903, Power: 1.1 kW, $Q_{max}$: 14 m$^3$/h (1)). Some of the power of the centrifugal pump dissipates into the turbulent flow, causing the rise of the temperature in the test loop. To provide a constant test loop water temperature, a bypass heat exchanger loop was installed into the facility. With the heat exchanger the water temperature was controlled and kept in 30 ±1 °C during the measurements. From the pump, the medium flows to a ball valve with a nominal diameter of initially ¾ inch (DN 32). The ball valve is not suitable for fine control of the mass flow and therefore it is followed by an angled seat valve. The fine control valve is followed by three HYDRUS ultrasonic flowmeter [2]. Multiple volumetric flow meters increase the accuracy of the volumetric flow measurement, which is very important for setting the inlet boundary condition for CFD calculations. The measuring channel section and the pump flow control subsystem are connected by a KPE pipe with an inner diameter of 26 mm. From here the water is fed through the diffuser cone to the flow straightening section. The flow straightener reduces disturbances caused by mechanical, flow control and pipe lining equipment. The 1 meter long seven-rod bundle was installed in the test channel section. A removable lid has been designed on the test section for ease of access which allows the change of the rod geometry. With this solution, different types of spacer grids and mixing vanes can be used.

A T-junction is placed after the test section; the medium discharges to the water tank trough other pipes lines. The schema of the test facility can be seen in the Figure 2 and the 3D model in Figure 1.

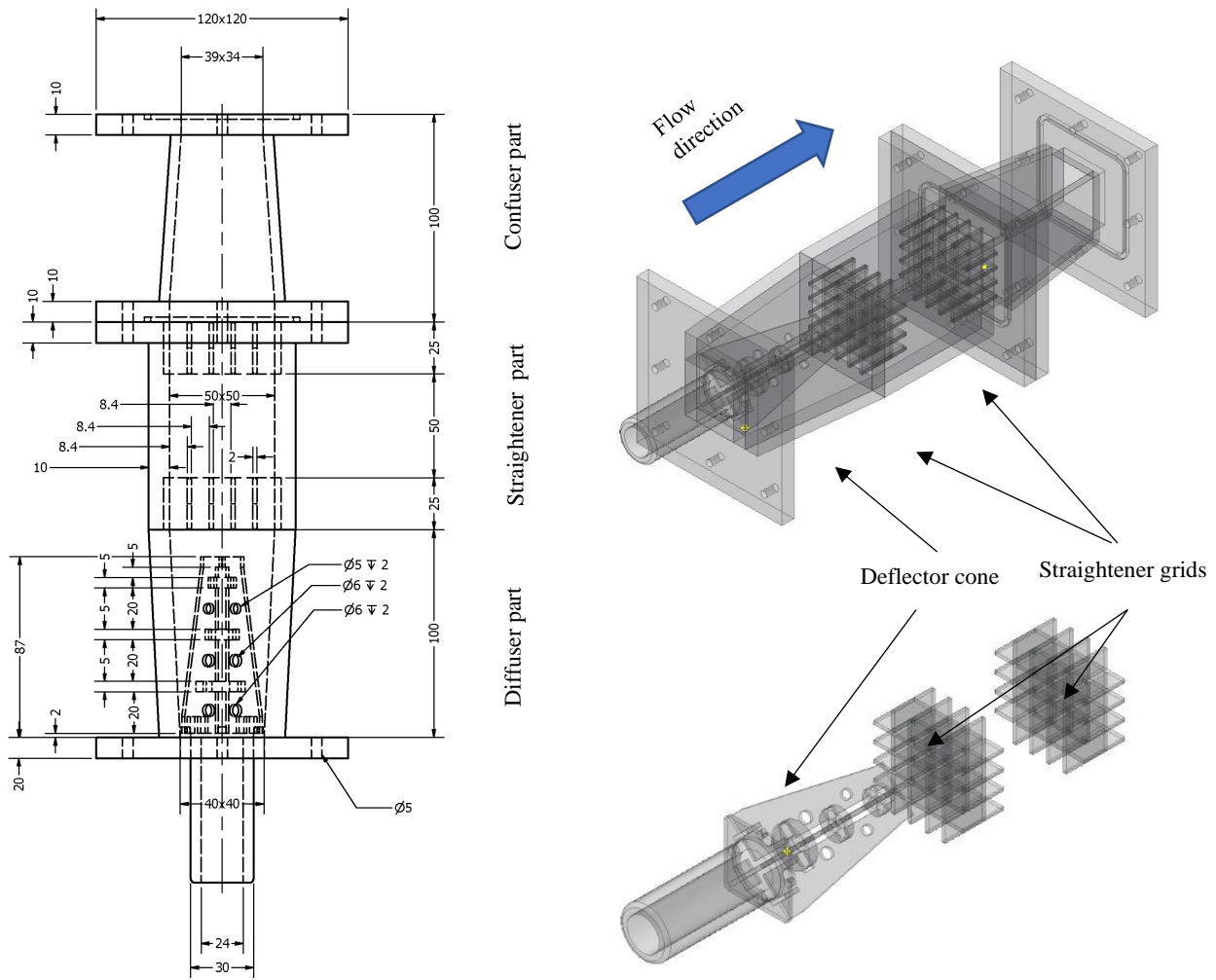

*Figure 3: 3D CAD model of the flow straightener*

The test section is designed to make the measurement section, the associated confuser and diffuser connectors easily interchangeable. The flow straightener includes two straightener grids and a deflector cone to restrict the flow from the effects of the pipe elbows upstream of the test-section. The cross section of the diffuser is 40x40 mm at the beginning and 50x50 mm at the end. A confuser adapter part is located above the straightener part. The cross section of the diffuser is 50x50 mm at the beginning and 39x34 mm at the end. At the exit of the confuser, the cross-section of the channel is the same as the cross-section of the measuring channel section. This avoids problems caused by sudden cross-section changes.

In the flow straightener, there is a part named deflector cone. This specifical part has the purpose of disrupting the main flow jet of the inlet pipe. It contains 4 perforated trapeze-shaped plates, which are perpendicular to each other. These plates hold the 3 deflector rings, which break the flow direction and eliminate the high-speed central jet. Due to the complexity of the geometry, the exact dimensions of these parts are not provided in written form but in 3D model format.

The middle section of the flow straightener contains two straightener grids with a height of 25 mm. The distance between the grids is 50 mm. The grids are made with compartment layout. The plates are made from 2 mm thick PMMA plates with equidistant grid spacing of 8.4 mm.

The cross-section of the measuring channel section exactly matches the hydraulic parameters of the reactor type under tests. Our 1 meter long bundle of rods is made of FEP (Fluorinated Ethylene Propylene) to meet the MIR criteria. The FEP polymer has a refractive index of 1.33 which is nearly the same as the refractive index of the working medium (water). The outer and inner diameter of the rod are 10/6 mm, and the inside of the rods is filled with ultrafiltered water. The rods are connected with small metal pins into the first and fourth spacer grids, and the spacer grids are connected to the channel wall with groove fitting.

## 2. PIV system

The PIV measurement system includes the following components:

- tracer particles: polyamide spheres with an average diameter of $d = 50$ µm (3),
- light source subsystem: Litron Nano L PIV dual Nd:YAG laser (maximum pulse energy: 135 mJ, wavelength: 532 nm, pulse length: ~6 ns, maximum flash frequency: 15 Hz) (4),
- beam guide arm and beam forming optics (5),
- image capture subsystem (camera): SpeedSense Lab 110 high-speed digital camera, resolution: 1 megapixel (1280x800), frame rate: 1630 fps, buffer: 12 GB (6),
- Synchronizer: Dantec Timer Box (80N77) (7),
- Synchronisation, image capture and processing software: Dantec DynamicStudio, latest stabile version 6.6 (8),
- camera and beam-optics positioning systems.

## 3. Measurement procedure

The flow straightener measurements were performed in the vertical measurement section. Figure 4 shows the schematic layout of the experiment. The illuminated volume is ~1.5 mm wide. We get information about the flow processes during the measurements from this volume. This measurement feature should also be considered during the CFD model result comparison. In the case of our current measurements, the illumination planes were positioned in the inside of the flow straightener section and other measurements were also taken at the beginning of the rod bundle test channel. The camera sees perpendicular to these planes.

Before starting the measurements, a so-called target sheet was placed in the appropriate position in the channel. The target sheet is a specially printed dotted sheet. Knowing the diameters of the dots and their positions helps to recover the real physical dimensions from the images. The coordinate axes can be identified using points of different diameters on the target. After a sufficient number of points have been detected, the conversion from pixel to millimetre distance is done automatically by a software (8).

In flow straightener measurements, 2000 image pairs were recorded in the vicinity of the straightener grids. Each gap of the compartment grid was examined. We aimed to observe the flow inside the flow volume. The first 100 image pair were discarded from the 2000 images captured because the lasers have a "warm-up" time requirement; therefore, the quality of the images at the beginning of the acquisition is not good.

To get a sufficiently detailed picture of the flow field, post-processing of the raw images is necessary. Figure 5 shows the steps of image processing. The first image shows the raw image (Figure 5/1). In the first step, an average image of 1900 image pairs was created (Figure 5/2). This average image was extracted from each image to reduce the effect of the elements that are present in each image (shadows, glitches and static elements) (Figure 5/3).

Laser light is not uniform in intensity along the length of the illuminated plane. Figure 5/4 shows an image processed by "image balancing" to correct for this unevenness of illumination. Since not all static elements can be eliminated from the images in this way, the static parts and regions not included in the flow field have to be masked out with digital masks. The row of Figure 5/5 shows the masked image, where only the polyamide particles that move with fluid are visible.

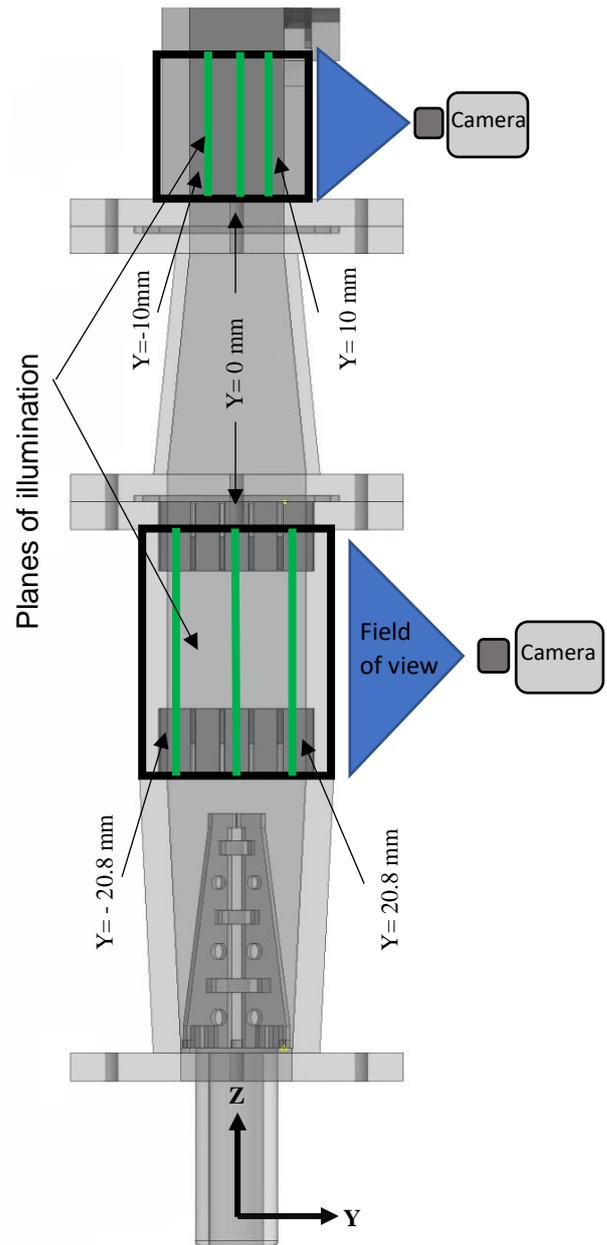

*Figure 4: Monitor planes*

After these steps, the individual image pairs were used to create the instantaneous vector fields separately. These vector fields show the chaotic velocity distribution typical of turbulent flow (Figure 5/6). From these 1900 vector diagrams, the time-averaged vector field describing the region after the spacer was created (Figure 5/7). With this method, not only the time-averaged velocities can be obtained, but also an estimate of the temporal fluctuations of the velocity vectors. In this way, we will not only be able to assign a vector value to a given pixel, but we will also be able to know its vector statistics.

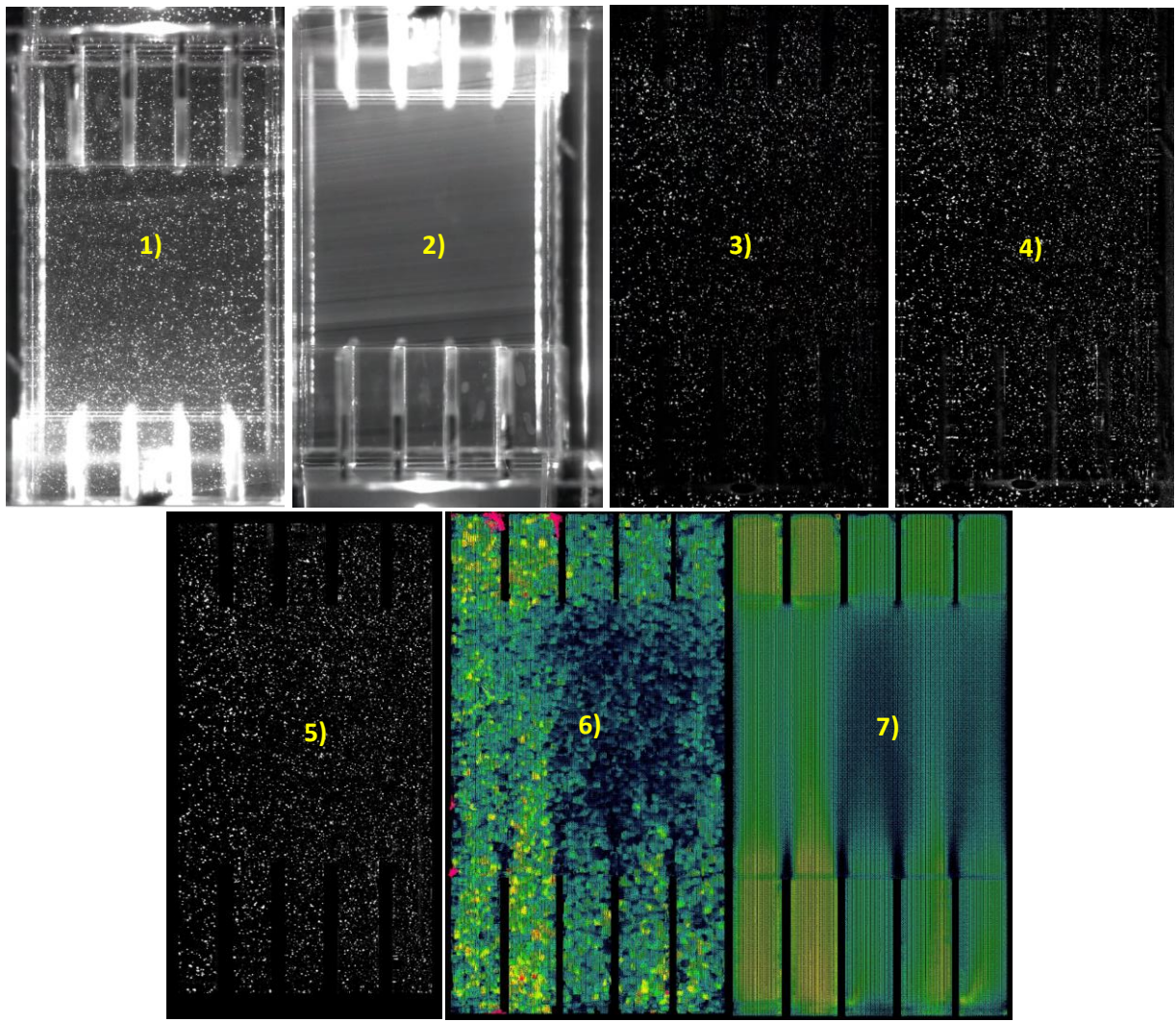

*Figure 5: The steps of the image post-processing*

## 4. Estimation of uncertainty in PIV measurements

In PIV measurement, the velocity of the particles flowing with the fluid is measured instead of the velocity of the flowing fluid. The density of the particles is approximately equal to the density of the liquid. The diameter of the particles in this case is 50 micrometres. Two digital images of the particle distribution are taken, from which the two-dimensional vector field can be calculated. The time interval between the two images can vary from a few microseconds to several milliseconds, depending on the velocity of the main flow.

In the interrogation areas, the velocity is assumed to be uniform during the image pair recording period. Knowing the delay between the recorded images and the displacement of the particles, the velocity vectors can be correlated to the interrogation regions using correlation methods (9).

Using calibration, the displacement (measured in pixels) can be converted to a metric value using the following formula (10):

$$u = \alpha \frac{\Delta X}{\Delta t} + \delta u \qquad (1)$$

Where: u is the physical velocity [m/s], α [m/pixel] is the conversion factor for magnification, ΔX [pixel] is the displacement of the recorded image, and Δt [s] is the time elapsed between the two images being recorded. The magnification factor α was determined by the calibration target. δu is difficult to detect systematically and is usually classified as an uncertainty factor rather than a measurement parameter.

In general, the measurement setup can be broken down into four subsystems:

- Calibration subsystem: converts the displacement in pixels into displacement in metric,
- Visualization: trace particles, illumination,
- Image recording: digital camera,
- Image processing: cross-correlation method, vector field calculation, etc.

The uncertainty in the target variables (flow velocities) is most affected by errors from the four subsystems.

| Main parameters | | Calibration | |
|---|---|---|---|
| Area investigated | 105 x 50 mm² | Calibration length on target $l_{sel}$ | 31.5 mm |
| Average flow velocity w | 2.0112 m/s | | |
| Flow cross section A | 2500 mm2 | Calibration length on the visualisation plane $L_{sel}$ | 365 pixel |
| Flow rate Q | 1.66667 l/s | Magnification $\alpha$ | 0.08630137 mm/pixel |
| **Flow visualisation** | | **Image recording** | |
| Trace particle | Polyamide spheres | Kamera | |
| Average diameter $d_p$ | 0.05 mm | Resolution | 1280 x 800 pixel |
| Average density | 1.02 g/cm³ | Frame rate | 1690 Hz |
| Light source | Litron Nano L PIV duál Nd:YAG laser | Objective | Nikon 60mm f/2.8 Micro-NIKKOR AF-D |
| Laser power | 138 mJ | Distance from the plane of illumination $l_t$ | 260 mm |
| Laser plane width | 1.5 mm | Angle of perspective $\varphi$ | 11.41 ° |
| Pulse frequency | 15 Hz | | |
| Time interval | 50 μs | | |
| **Data processing** | | | |
| Pixel value analysis | Cross correlation method | | |
| Interrogation area | 16 x16 pixel | | |
| Search area | 8 x 8 pixel | | |
| Sub-pixel analysis | three-point Gaussian fit | | |

*Table 1: Some basic data for the measurement system error calculation*

To achieve sufficiently accurate measurements, the estimates of random and systematic errors should be determined at the 95% confidence level and the resulting quadratic error function should be generated. This allows us to estimate the measurement uncertainty with 95% confidence.

Each element in equation (1) is subject to systematic error and random error, which introduce bias into the result and give the uncertainty of the measured value. Using the appropriate literature, a detailed uncertainty analysis was carried out which included the following sources (11) (10) (12) (13) (14) (15):

- Error sources and sensitivity factors for magnification $\alpha$
    - Reference length identification
    - Error caused by the image recording system
    - Error due to de-warping was neglected
- Error sources and sensitivity factors of $\Delta X$ image displacement
    - Error due to illumination
    - Error caused by the image recording system
    - Image processing, calculation of displacement
- Error sources and sensitivity factors of $\Delta t$ time delay
    - Error sources of the delay generator (timer) timing
    - Error sources of the laser pulse timing
- Error sources and sensitivity factors of $\delta u$ velocity difference
    - Flow following ability of the particles (trajectory)
    - Three-dimensional effects
    - Uncertainty due to volume flow adjustment
- The effect of sampling

At most points in the flow field, the error of our measurement is ~ 0.17 m/s. This relative error is naturally larger in the lower velocity sections (along walls), since most of the sources of error in the uncertainty analysis are constant, and few depend on the actual velocity vector of the measured flow. The uncertainty values fitted to the measurement points are included in the data series sent out.

The experimental data are available in .xls format and will be distributed directly to the participants. Please write an e-mail to **Gergely Imre Orosz <orosz@reak.bme.hu>**

## 5. Objective

The objective of this benchmark is the detailed investigation of the velocity field in the flow straightener present in the BME 7 pin ALLEGRO rod bundle test section. The goal is the comparison of the participants' results to test the different CFD codes, models and code applications (used meshes, turbulence models, difference schemes, user effects, etc.). Since PIV experiments of the mentioned flow straightener have been carried out at BME, comparisons with experimental data to validate the CFD codes are possible.

## 6. Input
### 6.1. Boundary conditions

The input data are coming from measurements of PIROUETTE test facility. The volumetric flow rate is set to 6 m³/h and is maintained under control thanks to the ultrasonic flow meters. The temperature of the water is 30°C and is maintained constant thanks to a heat exchanger, where at the second side there is mains tap water. Thanks to the heat exchanger, the model can be considered adiabatic in all its parts. The pressure of the water can be considered to be atmospheric at the outlet since the water is discharged in an open water tank. The properties of the water, such as density and dynamic viscosity, have to be computed considering the previously mentioned conditions, and the mass flow rate can be calculated considering these properties. The walls are considered smooth. The input data are summarized in Table 2. Regarding the physical walls, that include the straightener grids too, they are smooth and in no-slip condition. The flow in these conditions is turbulent both in the channels of the lamella that in the central part of the flow straightener:

$$Re_{centre} = \frac{4*\dot{m}}{\pi*D_h*\mu} = \frac{4*1.6594}{\pi*0.05*7.9735*10^{-4}} = 52996 \qquad (2)$$

$$Re_{channel} = \frac{4*\dot{m}}{\pi*D_h*\mu} = \frac{4*(1.6594\div25)}{\pi*8.4*10^{-3}*7.9735*10^{-4}} = 12618 \qquad (3)$$

|  | Volumetric flow rate [m³/h] | Temperature [°C] | Pressure [bar] | Density [kg/m³] | Dynamic viscosity [Pas] | Mass flow rate [kg/s] |
|---|---|---|---|---|---|---|
| Input value | 6 | 30 | 1 | 995,6515 | 7,9735 E-4 | 1,6594 |

*Table 2: Input data*

### 6.2. Geometry

In this section, details about the geometry can be found, in order to eliminate the errors that could come from a misunderstanding of the geometry and so from a wrong design. In Figure 6, the 3D ICEM geometry of the flow straightener can be seen, while in Figure 3 some details regarding the grid's dimensions are given.

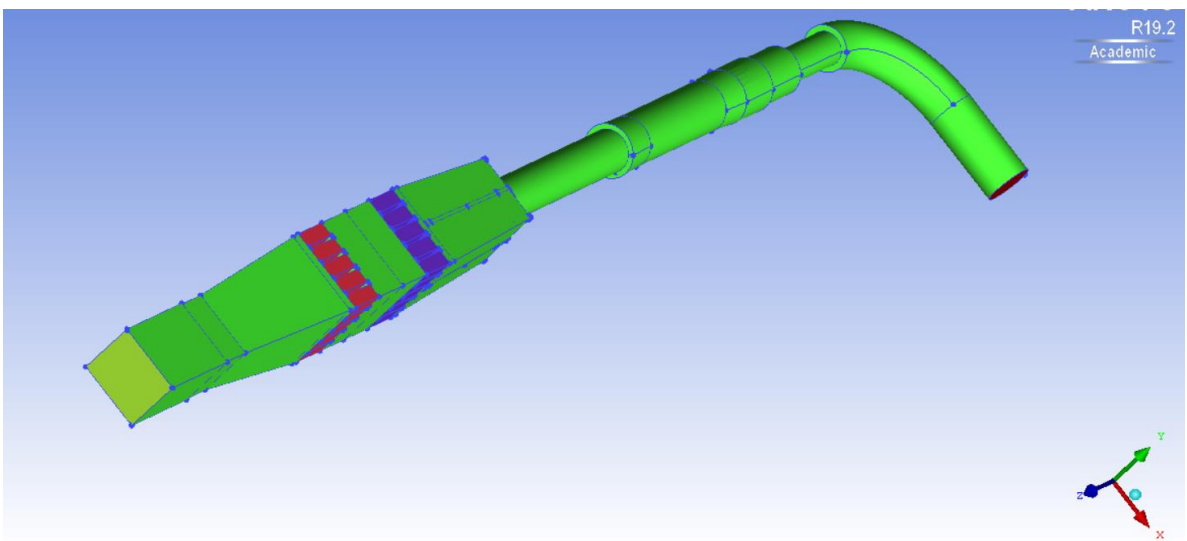

*Figure 6: Geometry of the model in ICEM CFD*

## 7. Hardware and Software Requirements: memory, files, approximated computational time

Hardware requirements strongly depend on the complexity of the applied model (resolution of the mesh, used turbulence model etc.). For the calculations, a 3D CFD code (e.g. CFX, FLUENT, STAR CD, etc.) is needed.

## 8. Output

### 8.1. Expected Results

- Axial (W in Z direction) and transversal (U in X direction) velocity at each monitoring line shown in Figure 7 for each of the planes shown in Figure 4. The planes shown in Figure 4 are the same as shown in Figure 7. The Y=0 mm plane being the center one. The y positions of the planes in the flow straightener section are: -20.8mm; 0; 20.8 mm and at the beginning of the test section are: -10 mm, 0 mm and 10 mm. Regarding the monitoring lines, considering the delivered geometry file, in Table 3 are reported the points from where to where all the center lines of the evaluation are going, so that the comparison of the results is made easier. The same points can be used for the other planes changing the y-direction value.
- The benchmark exercise will be organized in a semi-blind manner. The data of the beginning of the measuring test section part is provided openly before the exercise, and the measured velocity profiles inside the flow straightener will be used to rank the different modelling techniques (Figure 7).
- Model details: number of mesh nodes and elements, type of the mesh, turbulence model, boundary conditions, simulation type (steady state or transient), average value of Y+ (Yplus).
- It is important to note that the width of the monitor line is approximately 1,5 mm. For this reason, it is also recommended to perform the evaluation along a 1,5 mm wide strip (at the given Z height).

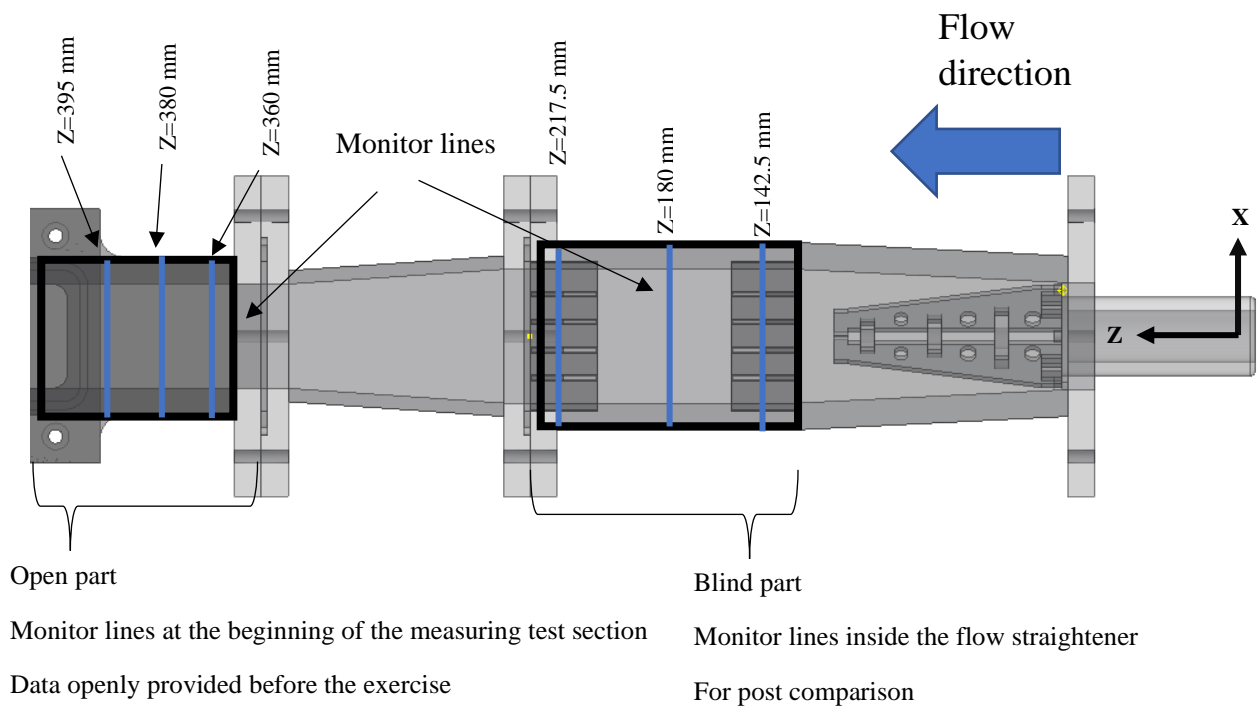

*Figure 7. Monitor lines*

|  | Starting point | | | Ending point | | |
|---|---|---|---|---|---|---|
|  | x | y | z | x | y | z |
| **Monitor lines inside the flow straightener (blind part)** | | | | | | |
| Y=0 | | | | | | |
| Z=142.5 | -25 | 0 | 142.5 | 25 | 0 | 142.5 |
| Z=180 | -25 | 0 | 180 | 25 | 0 | 180 |
| Z=217.5 | -25 | 0 | 217.5 | 25 | 0 | 217.5 |
| Y=-20.8 | | | | | | |
| Z=142.5 | -25 | -20.8 | 142.5 | 25 | -20.8 | 142.5 |
| Z=180 | -25 | -20.8 | 180 | 25 | -20.8 | 180 |
| Z=217.5 | -25 | -20.8 | 217.5 | 25 | -20.8 | 217.5 |
| Y=20.8 | | | | | | |
| Z=142.5 | -25 | 20.8 | 142.5 | 25 | 20.8 | 142.5 |
| Z=180 | -25 | 20.8 | 180 | 25 | 20.8 | 180 |
| Z=217.5 | -25 | 20.8 | 217.5 | 25 | 20.8 | 217.5 |
| **Monitor lines in the beginning of the measuring test section (open part)** | | | | | | |
| Y=0 | | | | | | |
| Z=360 | -20 | 0 | 360 | 20 | 0 | 360 |
| Z=380 | -20 | 0 | 380 | 20 | 0 | 380 |
| Z=395 | -20 | 0 | 395 | 20 | 0 | 395 |
| Y=-10 | | | | | | |
| Z=360 | -20 | -10 | 360 | 20 | -10 | 360 |
| Z=380 | -20 | -10 | 380 | 20 | -10 | 380 |
| Z=395 | -20 | -10 | 395 | 20 | -10 | 395 |
| Y=10 | | | | | | |
| Z=360 | -20 | 10 | 360 | 20 | 10 | 360 |
| Z=380 | -20 | 10 | 380 | 20 | 10 | 380 |
| Z=390 | -20 | 10 | 395 | 20 | 10 | 395 |

*Table 3: Coordinates of the starting point and the ending point of the monitoring lines, all the dimensions are in [mm] (Coordinate axis are shown in Figure 4. 6.- and 7)*

### 8.2. Files Format

The numerical data and the profiles are requested in MS Excel files.
The distributions should be printed in some kind of image files (e.g. *.jpg, *.png, *.bmp).
Proposed dimensions: velocity [m/s].

## 9. Outlook

Based on the experience gained during this benchmark a further benchmark exercise will be prepared in which the participants will investigate the flow field in a 7-rod bundle geometry with spacers relevant for ALLEGRO core geometries. The purpose of this section is to provide participants with appropriate practice for the second benchmark phase (7 rod bundle). Meanwhile, the results provided by their own models will be able to provide an appropriate inlet boundary condition for the second part.

## 10. Acknowledgment

This study received funding from EC in the frame of H2020 Grant Agreement No. 945041 (SafeG). The project was partially granted also by the Hungarian Atomic Energy Authority (Agreements No.: OAH-ABA-25/19-M, OAH-ABA-27/20-M, OAH-ABA-02/22-M).

# References

[1]     Data sheet: Economy MHIL 903,[Online], Available: https://www.pumps.co.za/PageFiles/5155384441.pdf.

[2]     Hydrus, DEIHL Metering, [Online], Available: https://www.bellflowsystems.co.uk/files/attachments/5084/HYDRUS.pdf.

[3]     D. Dynamics, Seeding particles for flow visualisation, LDA and PIV, Product information, Publication No.: Pi270003, 2002.

[4]     L. Lasers, Lamp Pumped lasers for PIV Applications from Litron, PB0101:3, 2010.

[5]     D. Dynamics, Light guide arm system, Publication No.: pi_257_v9, 2018.

[6]     V. Research, Phantom: Phantom Miro LAB/LC/R Series, ZDOC-64078-MA-0021 Rev 2, 2016.

[7]     D. Dynamics, Imaging Synchronization Devices, Product Information, Publication No.: pi:251_v6, 2011.

[8]     D. Dynamics, DynamicStudio – User's Guide, Build no.: 6.9.0059. Publication no.: 9040U1871, 2019.

[9]     M. Raffel and co-aouthors, Particle Image Velocimetry - A practical guide, Springer, Berlin, Germany, 2007.

[10]    The Visualization Society of Japan, Handbook of Particle Image Velocimetry, Morikita Publishing, 2002, p. Chapter 6: Assessment and management of measurement accuracy.

[11]    B. Yamaji, Thermal-hydraulics of a homogeneous molten salt fast reactor concept – experimental and numerical analyses, PhD thesis, Budapest university of technology and economics, Institute of Nuclear Techniques, 2016.

[12]    AANSI ASME PTC, Measurement Uncertainty, Supplement of Instrument and Apparatus, Part 1, New York: ASME, 1986.

[13]    Z. Szatmáry, Mérések kiértékelése, egyetemi jegyzet, (Evaluation of measurements, lecture textbook), Budapest: BME TTK, 2010.

[14]    W. G. Steele, R. A. Ferguson, R. P. Taylor and H. W. Coleman, Comparison of ANSI/ASME and ISO models for calculation of uncertainty, ISA Transactions 33, 1994 (339-352).

[15]    M. Shao, *Technical Issues for Narrow Angle Astrometry (STEP),* lecture note: https://www.ias.tsinghua.edu.cn/__local/A/B5/87/B94081E75AB6CF53678D0DC3BEE_5AB93B48_191E66.pdf?e=.pdf, 2021.10.18.

BME-NTI-1000/2023

# Preliminary Thesis on the Second Part of the ALLEGRO CFD Benchmark Exercise

## ALLEGRO CFD BENCHMARK

## PART 2

### Rod Bundle Benchmark description

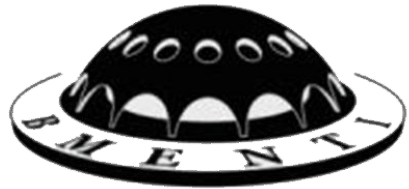

Institute of Nuclear Techniques (NTI)

Faculty of Natural Sciences (TTK)

BUDAPEST UNEIVERSITY OF TECHNOLOGY AND ECONOMICS (BME)

2023

| Name | Occupation | Institution | Contact |
|---|---|---|---|
| **Gergely Imre Orosz** | **PhD student** | **BME NTI** | **orosz@reak.bme.hu** |
| Mathias Peiretti | MSc student | BME NTI/ Politecnico di Torino | matthias.peiretti@edu.bme.hu |
| Boglárka Magyar | BSc student | BME NTI | bogi0614@gmail.com |
| Dániel Szerbák | MSc student | BME NTI | szerbakdani@gmail.com |
| Dániel Kacz | PhD student | BME NTI | kacz.daniel@reak.bme.hu |
| Béla Kiss | Research assistant | BME NTI | kiss@reak.bme.hu |
| Gábor Zsíros | Research assistant | BME NTI | zsiros@reak.bme.hu |
| Prof. Attila Aszódi | Professor | BME NTI | aszodi@reak.bme.hu |

Budapest, Hungary, 2023.02.01

## 1. Short description

In the first part of the SafeG benchmark exercise, the flow field of the flow straightener part of the PIROUETTE experimental facility was investigated. At BME (Budapesti Műszaki és Gazdaságtudományi Egyetem - Budapest University of Technology and Economics) NTI (Nukleáris Technikai Intézet - Institute of Nuclear Techniques), a 7 pin ALLEGRO rod bundle test facility has been built in order to investigate the hydraulic behavior of the coolant in such design and to develop CFD models that could properly simulate the flow conditions in the ALLEGRO core. PIROUETTE (PIv ROd bUndlE Test faciliTy at bmE) is a test facility, which was designed to investigate the emerging flow conditions in various nuclear fuel assembly rod bundles. The measurement method is based on Particle Image Velocimetry (PIV) with Matching of Index of Refractory (MIR) method. Despite the coolant will be helium in ALLEGRO, the working fluid is water in the measurements, but both fluids are in the same Reynolds-number range.

In the test loop, it was necessary to install a flow straightener that was able to condition the velocity field before the rod bundle. In the benchmark exercise Part1, the conditions in this flow straightener part are simulated. The results of CFD simulations could be used to improve the understanding of the inlet conditions in the rod bundle test section.

The herein proposed benchmark deals with the 3D CFD modeling of the velocity field within the 7 pin rod bundle placed in the test section. The geometry of the test section will be given to the participants in an easy-to-use 3D format (.obj, .stp or .stl). Link for ancillary files:

https://drive.google.com/drive/folders/11MEuNLHhmpW8Cr9DUCx5s1JjtgdtgMGk?usp=sharing

In Figure 1, an overview of the test loop is provided. It can be seen where the vertical test section was placed.

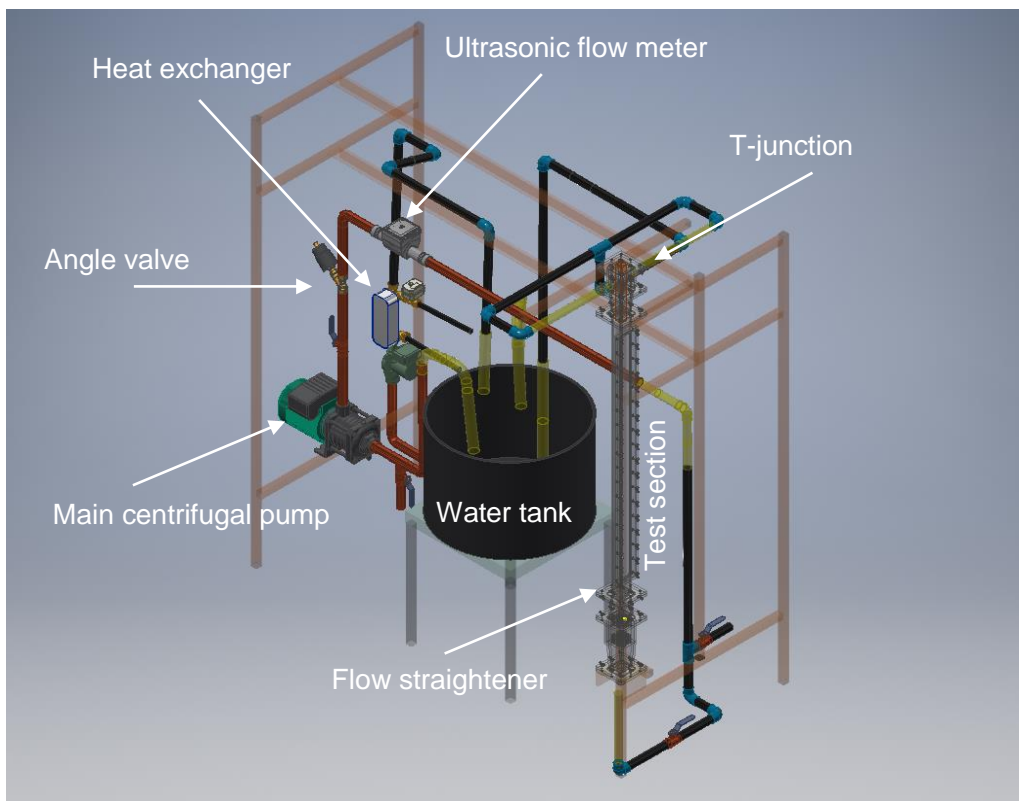

*Figure 1: The structure of the PIROUETTE facility*

## 1.1. Test section

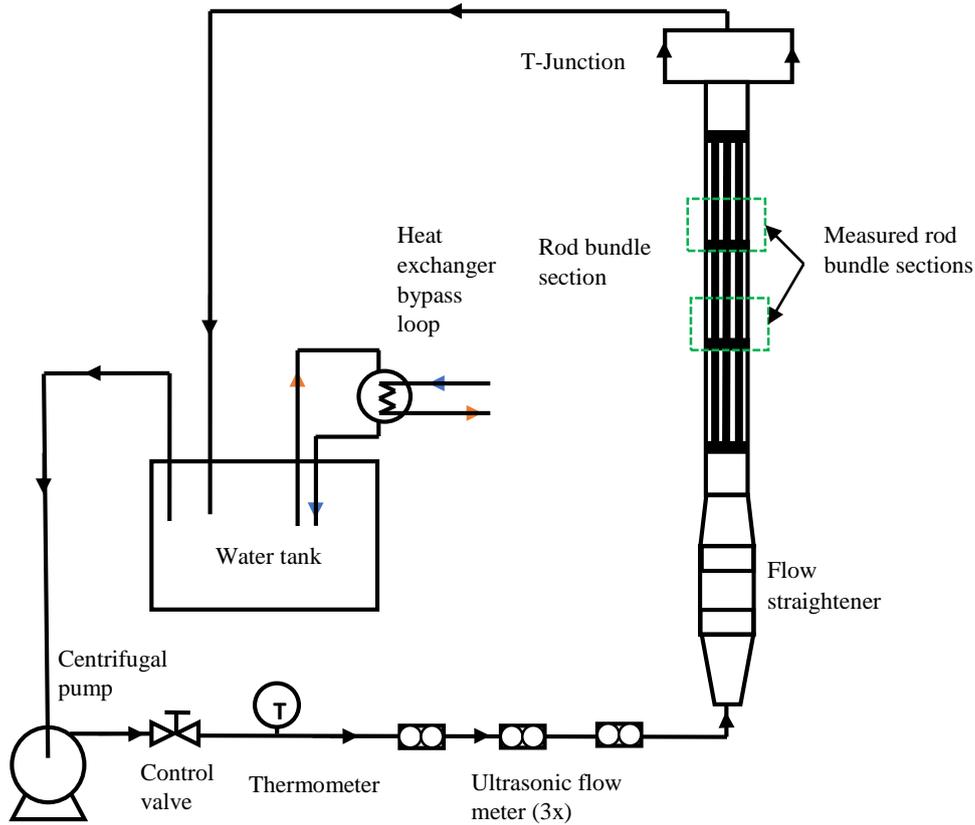

*Figure 2: The schematic of the PIROUETTE facility*

The structure and the main parts of the test facility can be seen in Figure 1. The installation contains a 1-meter-long vertically arranged seven pin rod bundle in the test section. The water flow is provided by the main centrifugal pump (Type: Wilo MHIL 903, Power: 1.1 kW, $Q_{max}$: 14 m$^3$/h [1]). Some of the power of the centrifugal pump dissipates into the turbulent flow, causing the rise of the temperature in the test loop. To provide a constant test loop water temperature, a bypass heat exchanger loop was installed into the facility. With the heat exchanger the water temperature was controlled and kept in 30 ±1 °C during the measurements. From the pump, the medium flows to a ball valve with a nominal diameter of initially ¾ inch (DN 32). The ball valve is not suitable for fine control of the mass flow and therefore it is followed by an angled seat valve. The fine control valve is followed by three HYDRUS ultrasonic flowmeter [2]. Multiple volumetric flow meters increase the accuracy of the volumetric flow measurement, which is very important for setting the inlet boundary condition for CFD calculations. The measuring channel section and the pump flow control subsystem are connected by a KPE pipe with an inner diameter of 26 mm. From here the water is fed through the diffuser cone to the flow straightening section. The flow straightener reduces disturbances caused by mechanical, flow control and pipe lining equipment. The 1-meter-long seven-rod bundle was installed in the test channel section. A removable lid has been designed on the test section for ease of access which allows the change of the rod geometry. A T-junction is placed after the test section; the medium discharges to the water tank trough other pipes lines. The schema of the test facility can be seen in Figure 2 and the 3D model in Figure 1. The detailed compilation of the test section is in Figure 3.

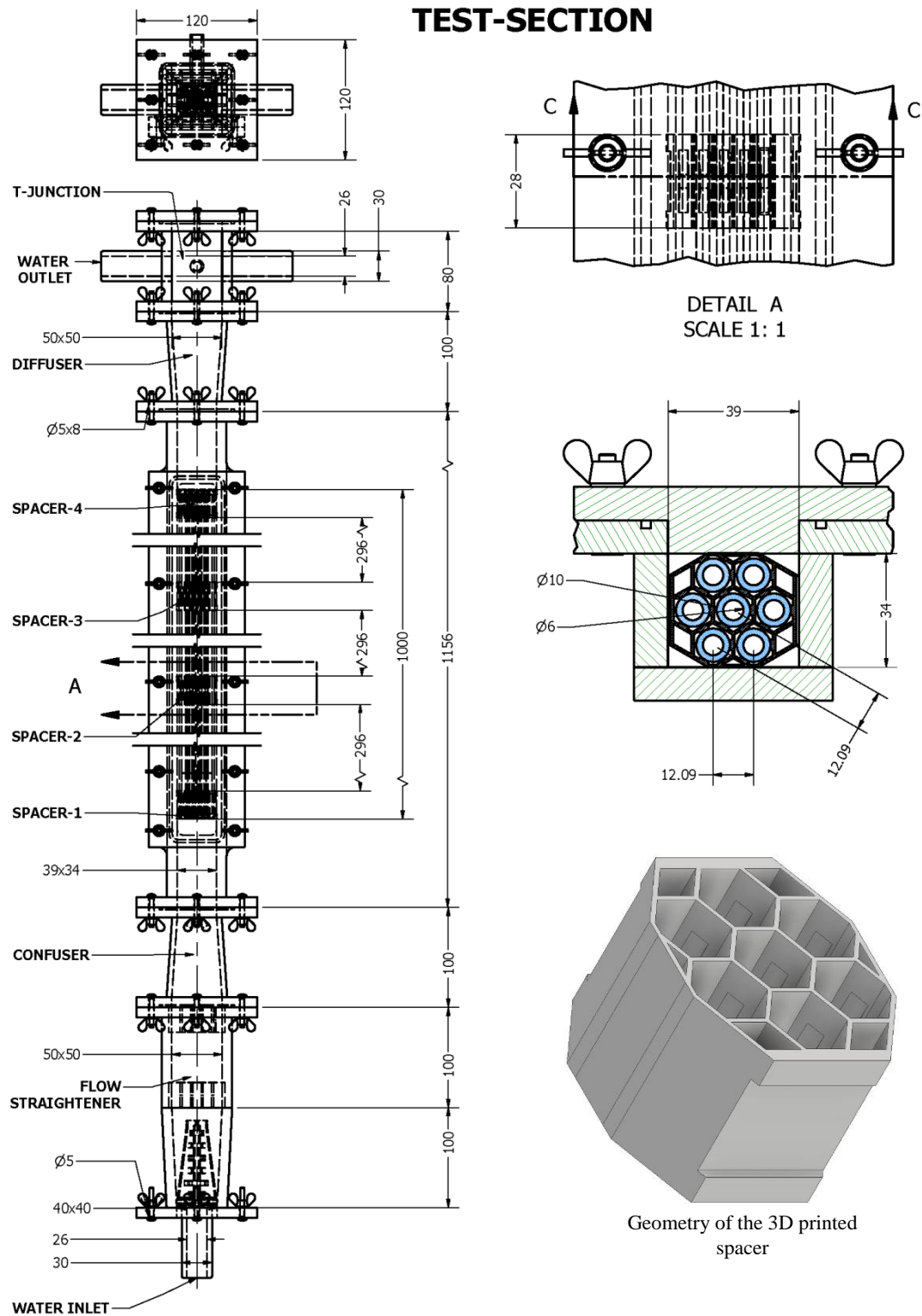

*Figure 3: Compilation of the test section*

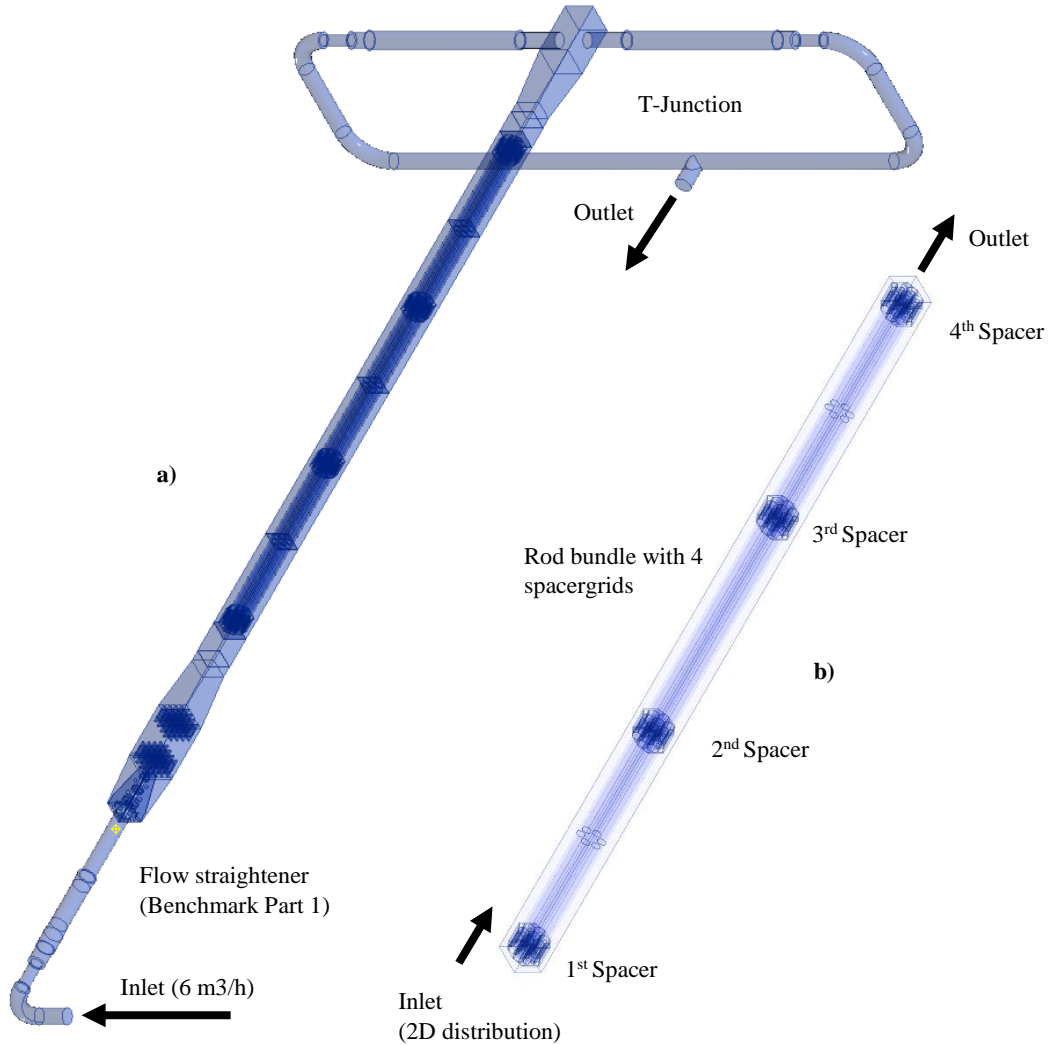

*Figure 4:* 3D CAD model of the test section of PIROUETTE facility (a) and the included rod bundle (b)

The test section is designed to includes the measurement section, the associated confuser, and diffuser connectors. These parts are easily interchangeable to perform measurements for different nuclear reactor rod bundles. The cross-section of the measuring channel is a little bigger than the ALLEGRO, but on the basis of the similarity theory [2], the hydraulic parameters are exactly the same. The 1-meter-long test bundle is made of FEP (Fluorinated Ethylene Propylene) to meet the refractory index of water. The FEP polymer has a refractory index of 1.33 which is nearly the same as of the working fluid (water). The outer and inner diameter of the rods are 10/6 mm, and the inside of the rods was filled with ultrafiltrated water. The diameter of the ALLEGRO refractory fuel rod is 9.1 mm. Therefore, the upscaling ratio of the channel was 10/9.1=1.0989. The rods are connected with pins into the first and fourth spacer grids, and the spacer grids are connected to the wall of the channel with groove fitting.

The distance between the spacers is 296 mm in axial direction. The spacers were designed according to the ALLEGRO GFR assembly spacers, with slight modifications to fit the measurement requirements. The wall thickness of the spacers is 0.8 mm. The grids were made by a high resolution 3D printing method (SLA) with special rigid composite resin (see in Figure 3).

## 2. PIV system

The PIV measurement system includes the following components:

- tracer particles: polyamide spheres with an average diameter of d = 50 µm [3],
- light source subsystem: Litron Nano L PIV dual Nd:YAG laser (maximum pulse energy: 135 mJ, wavelength: 532 nm, pulse length: ~6 ns, maximum flash frequency: 15 Hz) [4],
- beam guide arm and beam forming optics [5],
- image capture subsystem (camera): SpeedSense Lab 110 high-speed digital camera, resolution: 1 megapixel (1280x800), frame rate: 1630 fps, buffer: 12 GB [6],
- Synchronizer: Dantec Timer Box (80N77) [7],
- Synchronisation, image capture and processing software: Dantec DynamicStudio, latest stabile version 6.6 [8],
- camera and beam-optics positioning systems.

## 3. Measurement procedure

The rod bundle measurements were performed in the vertical measurement section. Figure 5 shows the schematic layout of the experiment. The illuminated volume is ~1.5 mm wide. We get information about the flow processes during the measurements from this volume. This measurement feature should also be considered during the CFD model result comparison. In the case of our current measurements, the illumination planes were positioned to intersect the two outer rod. The camera sees perpendicular to this plane.

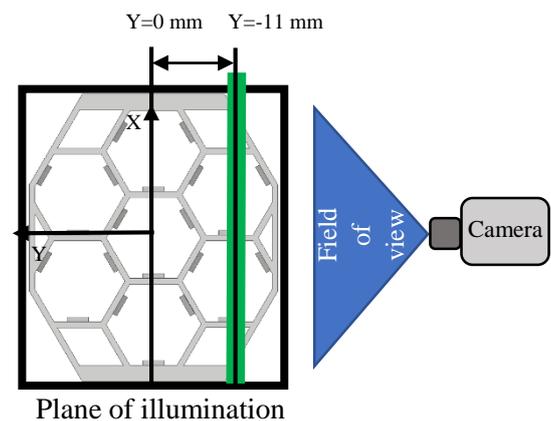

*Figure 5: The schematic layout of the investigated planes*

With a so-called target sheet, a measurement calibration has been made. With the calibration, the conversion from pixel to millimeter distance is done automatically by a software [8].

In the rod bundle measurements, 2000 image pairs were recorded in the vicinity of the second and third spacer grids. During the benchmark, the pictures captured behind the third grid will be used. The flow field was captured on a plane positioned at Y=-11 mm (Figure 5). The plane passes through the side channels of the test section, the two outer rods, and the subchannel between the two outer rods.

The first 100 image pair were discarded from the 2000 images captured because the lasers have a "warm-up" time requirement; therefore, the quality of the images at the beginning of the acquisition is not good.

To get a sufficiently detailed picture of the flow field, post-processing of the raw images is necessary. Figure 6 shows the steps of image processing. The first image shows the raw image (Figure 6/1). In the first step, an average image of 1900 image pairs was created (Figure 6/2). This average image was extracted from each image to reduce the effect of the elements that are present in each image (shadows, glitches and static elements) (Figure 6/3).

Laser light is not uniform in intensity along the length of the illuminated plane. Figure 6/4 shows an image processed by "image balancing" to correct for this unevenness of illumination. Since not all static elements can be eliminated from the images in this way, the static parts and regions not included in the flow field have to be masked out with digital masks. The row of Figure 6/5 shows the masked image, where only the polyamide particles that move with fluid are visible.

After these steps, the individual image pairs were used to create the instantaneous vector fields separately. These vector fields show the chaotic velocity distribution typical of turbulent flow (Figure 6/6). From these 1900 vector diagrams, the time-averaged vector field describing the region after the spacer was created (Figure 6/7). With this method, not only the time-averaged velocities can be obtained, but also an estimate of the temporal fluctuations of the velocity vectors. In this way, we will not only be able to assign a vector value to a given pixel, but we will also be able to know its vector statistics.

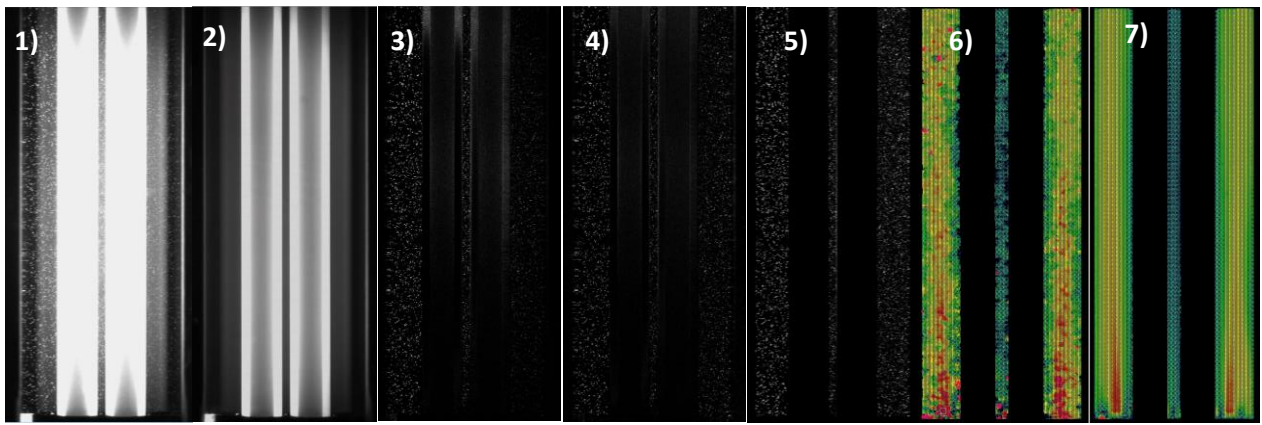

*Figure 6: The steps of the image post-processing in the rod bundle*

### 4. Estimation of uncertainty in PIV measurements

In PIV measurement, the velocity of the particles flowing with the fluid is measured instead of the velocity of the flowing fluid. The density of the particles is approximately equal to the density of the liquid. The diameter of the particles in this case is 50 micrometers. Two digital images of the particle distribution are taken, from which the two-dimensional vector field can be calculated. The time interval between the two images can vary from a few microseconds to several milliseconds, depending on the velocity of the main flow.

In the interrogation areas, the velocity is assumed to be uniform during the image pair recording period. Knowing the delay between the recorded images and the displacement of the particles, the velocity vectors can be correlated to the interrogation regions using correlation methods [9].

Using calibration, the displacement (measured in pixels) can be converted to a metric value using the following formula [10]:

$$u = \alpha \frac{\Delta X}{\Delta t} + \delta u \qquad (1)$$

Where: u is the physical velocity [m/s], α [m/pixel] is the conversion factor for magnification, ΔX [pixel] is the displacement of the recorded image, and Δt [s] is the time elapsed between the two images being recorded. The magnification factor α was determined by the calibration target. δu is difficult to detect systematically and is usually classified as an uncertainty factor rather than a measurement parameter.

In general, the measurement setup can be broken down into four subsystems:

- Calibration subsystem: converts the displacement in pixels into displacement in metric,
- Visualization: trace particles, illumination,
- Image recording: digital camera,
- Image processing: cross-correlation method, vector field calculation, etc.

The uncertainty in the target variables (flow velocities) is most affected by errors from the four subsystems.

| Main parameters | | Calibration | |
|---|---|---|---|
| Area investigated | 105 x 50 mm$^2$ | Calibration length on target $l_{sel}$ | 31.5 mm |
| Average flow velocity w | 2.0112 m/s | | |
| Flow cross section A | 828.7 mm$^2$ | Calibration length on the visualization plane $L_{sel}$ | 365 pixel |
| Flow rate Q | 1.66667 l/s | Magnification $\alpha$ | 0.08630137 mm/pixel |
| Flow visualization | | Image recording | |
| Trace particle | Polyamide spheres | Camera | |
| Average diameter $d_P$ | 0.05 mm | Resolution | 1280 x 800 pixel |
| Average density | 1.02 g/cm$^3$ | Frame rate | 1690 Hz |
| Light source | Litron Nano L PIV duál Nd:YAG laser | Objective | Nikon 60mm f/2.8 Micro-NIKKOR AF-D |
| Laser power | 138 mJ | Distance from the plane of illumination $l_t$ | 260 mm |
| Laser plane width | 1.5 mm | Angle of perspective $\varphi$ | 11.41 ° |
| Pulse frequency | 15 Hz | | |
| Time interval | 50 μs | | |
| Data processing | | | |
| Pixel value analysis | Cross correlation method | | |
| Interrogation area | 16 x16 pixel | | |
| Search area | 8 x 8 pixel | | |
| Sub-pixel analysis | three-point Gaussian fit | | |

*Table 1: Some basic data for the measurement system error calculation*

To achieve sufficiently accurate measurements, the estimates of random and systematic errors should be determined at the 95% confidence level and the resulting quadratic error function should be generated. This allows us to estimate the measurement uncertainty with 95% confidence.

Each element in equation (1) is subject to systematic error and random error, which introduce bias into the result and give the uncertainty of the measured value.

Using the appropriate literature, a detailed uncertainty analysis was carried out which included the following sources [11] [10] [12] [13] [14] [15]:

- Error sources and sensitivity factors for magnification $\alpha$
    - Reference length identification
    - Error caused by the image recording system
    - Error due to de-warping was neglected
- Error sources and sensitivity factors of $\Delta X$ image displacement
    - Error due to illumination
    - Error caused by the image recording system
    - Image processing, calculation of displacement
- Error sources and sensitivity factors of $\Delta t$ time delay
    - Error sources of the delay generator (timer) timing
    - Error sources of the laser pulse timing
- Error sources and sensitivity factors of $\delta u$ velocity difference
    - Flow following ability of the particles (trajectory)
    - Three-dimensional effects
    - Uncertainty due to volume flow adjustment
- The effect of sampling

In the respect of the average flow velocity in the rod bundle (2.0112 m/s, see in Table 1), the error of our measurement is ~ 0.22 m/s (~10.58% of the average velocity). This relative error is naturally larger in the lower velocity sections (along walls), since most of the sources of error in the uncertainty analysis are constant, and few depend on the actual velocity vector of the measured flow. The uncertainty values fitted to the measurement points are included in the data series sent out.

The experimental data are available in .xls format and will be distributed directly to the participants. Please write an e-mail to **Gergely Imre Orosz <orosz@reak.bme.hu>**

## 5. Objective

The objective of this benchmark is the detailed investigation of the velocity field in the 7 pin ALLEGRO rod bundle test section. The goal is the comparison of the participants' results to test the different CFD codes, models and code applications (used meshes, turbulence models, difference schemes, user effects, etc.). Since PIV experiment of the mentioned BME 7 pin ALLEGRO rod bundle test section has been carried out at BME, comparisons with experimental data to validate the CFD codes are possible.

## 6. Input
### 6.1. Boundary conditions

The input data are coming from measurements of PIROUETTE test facility. The volumetric flow rate is set to 6 m$^3$/h and is maintained under control thanks to the ultrasonic flow meters. The temperature of the water is 30°C and is maintained constant thanks to a heat exchanger, where at the second side there is mains tap water. Thanks to the heat exchanger, the model can be considered adiabatic in all its parts. The pressure of the water can be considered to be atmospheric at the outlet since the water is discharged in an open water tank.

The properties of the water, such as density and dynamic viscosity, have to be computed considering the previously mentioned conditions, and the mass flow rate can be calculated considering these properties. The walls are considered smooth. The input data are summarized in Table 2.

The calculated Re-number for the rod bundle:

$$D_{Hydraulic\_Bundle} = \frac{4*\dot{A}}{K} = \frac{4*0.000827971}{0.368573} = 0.008983 \text{ m} \quad (2)$$

$$Re_{bundle} = \frac{W_{average}*D_{Hydraulic\_Bumdle}*\rho}{\mu} = \frac{2.0112*0.008983*995.6515}{7.9735*10^{-4}} = 22\,555 \quad (3)$$

|  | Volumetric flow rate [m$^3$/h] | Temperature [°C] | Pressure [bar] | Density [kg/m$^3$] | Dynamic viscosity [Pas] | Mass flow rate [kg/s] |
|---|---|---|---|---|---|---|
| Input value | 6 | 30 | 1 | 995.6515 | 7.9735 E-4 | 1.6594 |

*Table 2: Input data*

2-dimensional inlet velocity distribution is provided to the shorter rod bundle domain. The shorter rod bundle domain is presented Figure 4/b. The inlet velocity distribution file name is ROD_BUNDLE_INLET_VELOCITY.csv. The file contains the velocity, turbulent kinetic energy and turbulent kinetic energy dissipation distributions.

### 6.2. Geometry

In this section, details about the geometry can be found in order to eliminate the errors that could come from a misunderstanding of the geometry and so from a wrong design. The 3D CAD geometry of the rod bundle can be seen in Figure 4. The following .stl files will be delivered to the benchmark participants: COMPLETE_TEST_SECTION.stl /.obj /.step. and ROD_BUNDLE.stl /.obj /.step.

- The COMPLETE_TEST_SECTION.stl file involves the fluid domain of the complete rod bundle, with the flow straightener and the upper T-junction sections (see in Figure 4/a).The reason we provide the complete geometry is to give users the freedom to customize the fluid domain they want to describe.
- The ROD_BUNDLE.stl file is describing the rob bundle fluid section and the inlet velocity profile is provided for this shorter model. The inlet velocity distribution file name is ROD_BUNDLE_INLET_VELOCITY.csv

### 6.3. Hardware and Software Requirements: memory, files, approximated computational time

Hardware requirements strongly depend on the complexity of the applied model (resolution of the mesh, used turbulence model etc.). For the calculations, a 3D CFD code (e.g. CFX, FLUENT, STAR CD, etc.) is needed.

## 7. Output
### 7.1. Expected Results
- Axial (W in Z direction) and transversal (U in X direction) velocity components at each monitoring line (including coordinates) shown in Figure 7 for the green plane shown in Figure 5. The plane shown in Figure 5 is the same as shown in Figure 7. The y position of the plane is -11 mm. The monitor lines are located behind the 3rd spacer grid in different distances from the end of the grid. The first monitor line is located 0.5D (5 mm) behind the spacer, where D is the diameter of the rods (10 mm). Regarding the monitoring lines, considering the delivered geometry file, in Table 3 are reported the points from where to where all the lines of the evaluation are going, so that the comparison of the results is made easier. The same points can be used for the other lines changing the z-direction value.
- In addition, results are also evaluated at the midline of the gap between the two outer rods. The coordinates of this center line are also given in Table 3.
- Model details: number of mesh nodes and elements, type of the mesh, turbulence model, boundary conditions, simulation type (steady state or transient), average value of Y+ (Yplus).
- It is important to note that the width of the laser sheet is approximately 1,5 mm. For this reason, it is also recommended to perform the evaluation for a 1,5 mm wide region along the monitor line (at the given Z height).

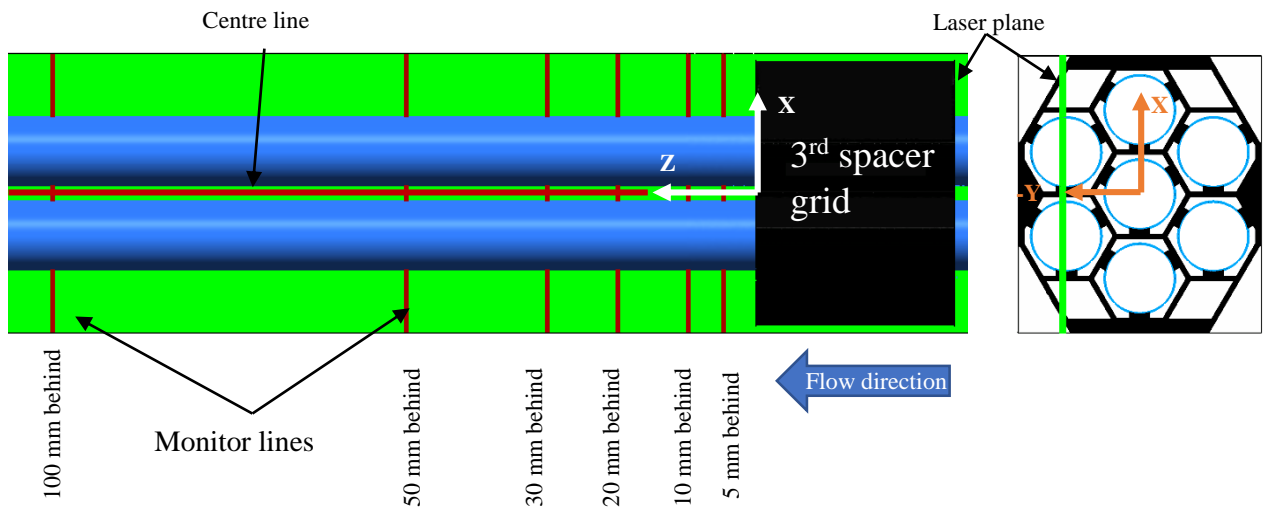

Figure 7. Monitor lines *view* from -Y direction

|  | Starting point | | | Ending point | | |
|---|---|---|---|---|---|---|
|  | x | y | z | x | y | z |
| **Monitor lines behind the 3rd Spacer** | | | | | | |
| Y=-11 | | | | | | |
| 0.5D | -20 | -11 | 1098 | 20 | -11 | 1098 |
| 1D | -20 | -11 | 1103 | 20 | -11 | 1103 |
| 2D | -20 | -11 | 1113 | 20 | -11 | 1113 |
| 3D | -20 | -11 | 1123 | 20 | -11 | 1123 |
| 5D | -20 | -11 | 1143 | 20 | -11 | 1143 |
| 10D | -20 | -11 | 1193 | 20 | -11 | 1193 |
| X=0 | | | | | | |
| Center line | 0 | -11 | 1093 | 0 | -11 | 1193 |

*Table 3: Coordinates of the starting point and the ending point of the monitoring lines, all the dimensions are in [mm] (Coordinate axis are shown in Figure 5. and 7.)*

### 7.2. Files Format

The numerical data and the profiles are requested in MS Excel files.
The distributions should be printed in some kind of image files (e.g. *.jpg, *.png, *.bmp).
Proposed dimensions: velocity [m/s].

# References

[1] Data sheet: Economy MHIL 903,[Online], Available: https://www.pumps.co.za/PageFiles/5155384441.pdf.

[2] Hydrus, DEIHL Metering, [Online], Available: https://www.bellflowsystems.co.uk/files/attachments/5084/HYDRUS.pdf.

[3] Dantec Dynamics, Seeding particles for flow visualisation, LDA and PIV, Product information, Publication No.: Pi270003, 2002.

[4] Litron Lasers, Lamp Pumped lasers for PIV Applications from Litron, PB0101:3, 2010.

[5] Dantec Dynamics, Light guide arm system, Publication No.: pi_257_v9, 2018.

[6] Vision Research, Phantom: Phantom Miro LAB/LC/R Series, ZDOC-64078-MA-0021 Rev 2, 2016.

[7] Dantec Dynamics, Imaging Synchronization Devices, Product Information, Publication No.: pi:251_v6, 2011.

[8] Dantec Dynamics, DynamicStudio – User's Guide, Build no.: 6.9.0059. Publication no.: 9040U1871, 2019.

[9] M. Raffel and co-aouthors, Particle Image Velocimetry - A practical guide, Springer, Berlin, Germany, 2007.

[10] The Visualization Society of Japan, Handbook of Particle Image Velocimetry, Morikita Publishing, 2002, p. Chapter 6: Assessment and management of measurement accuracy.

[11] B. Yamaji, Thermal-hydraulics of a homogeneous molten salt fast reactor concept – experimental and numerical analyses, PhD thesis, Budapest university of technology and economics, Institute of Nuclear Techniques, 2016.

[12] AANSI ASME PTC, Measurement Uncertainty, Supplement of Instrument and Apparatus, Part 1, New York: ASME, 1986.

[13] Z. Szatmáry, Mérések kiértékelése, egyetemi jegyzet, (Evaluation of measurements, lecture textbook), Budapest: BME TTK, 2010.

[14] W. G. Steele, R. A. Ferguson, R. P. Taylor and H. W. Coleman, Comparison of ANSI/ASME and ISO models for calculation of uncertainty, ISA Transactions 33, 1994 (339-352).

[15] M. Shao, *Technical Issues for Narrow Angle Astrometry (STEP),* lecture note: https://www.ias.tsinghua.edu.cn/__local/A/B5/87/B94081E75AB6CF53678D0DC3BEE_5AB93B48_191E66.pdf?e=.pdf, 2021.10.18.